\documentclass[aps,prl,twocolumn,superscriptaddress,nofootinbib,longbibliography]{revtex4-2}
\usepackage[colorlinks]{hyperref}   
\usepackage{amsmath,amssymb,graphicx,bm}
\usepackage{braket}
\usepackage{comment}
\usepackage{xcolor}

\begin{document}
	
	\title{Eigenstate-assisted realization of general quantum controlled unitaries with a fixed cost}
	
	\author{Carlos Navas-Merlo}
	\thanks{Corresponding author: \href{mailto:carlos.navas@uva.es}{carlos.navas@uva.es}}
	\affiliation{Departamento de Teoría de la Señal y Comunicaciones e Ingeniería Telemática, Universidad de Valladolid, Paseo Bel\'en 15, 47011 Valladolid, Spain}
	
	\author{Juan Carlos García Escartín}
	\email{juagar@uva.es}
	\affiliation{Departamento de Teoría de la Señal y Comunicaciones e Ingeniería Telemática, Universidad de Valladolid, Paseo Bel\'en 15, 47011 Valladolid, Spain}
	\affiliation{Laboratory for Disruptive Interdisciplinary Science (LaDIS), Universidad de Valladolid, 47011 Valladolid, Spain}
	
	\date{\today}

	\begin{abstract}
		Controlled unitary gates are a basic element in many quantum algorithms. Converting a general unitary $U$ with a known decomposition into its controlled version, controlled-$U$, can introduce a large overhead in terms of the depth of the circuit. We present a general method to take any unitary $U$ into controlled-$U$ using a fixed circuit with 4 CNOT gates and 2 Toffoli gates per qubit. For $n$-qubit unitaries and one control qubit, we require $2n+1$ qubits and a circuit that can generate an eigenstate of $U$, for which there are many cost-effective known algorithms. The method also works for any black block implementation of $U$, achieving a constant-depth realization independent of its decomposition. We illustrate its use in the Hadamard test and discuss applications to variational and quantum machine-learning algorithms.
	\end{abstract}
	
	\maketitle
	
	\textit{Introduction.—}
	Quantum computation is rapidly becoming a practical reality, with current technology already enabling noisy intermediate-scale quantum (NISQ) processors~\cite{Preskill2018,Arute2019,Nam2020}. Although these devices can perform coherent operations with tens or hundreds of qubits, they remain too noisy to execute deep circuits such as those required for Shor’s factoring algorithm or large-scale error-corrected computations~\cite{Shor1997,Gottesman2002}. Consequently, there is an intense search for algorithms that can extract quantum advantage using shallow circuits, most notably in areas such as molecular simulation~\cite{Kandala2017,Cao2019}, quantum machine learning~\cite{Schuld2018,Dunjko2018}, and quantum metrology~\cite{Giovannetti2011}. Because noise and decoherence scale with circuit depth, any algorithm aiming to achieve practical impact in the NISQ era must minimize gate overhead. For that reason, numerous works have focused on optimizing gate counts and performing depth-efficient compilation of unitaries~\cite{Maslov2017,Bocharov2013,Cross2019}.\\
	
	As device fidelities improve, the dominant overhead increasingly stems not from single- or two-qubit operations themselves but from composite primitives such as controlled unitaries. Implementing a controlled-$U$ for an $n$-qubit unitary $U$ of depth $d_U$ generally requires $\mathcal{O}(n\,d_U)$ two-qubit gates and the promotion of each subgate to a controlled counterpart, which substantially increases circuit depth and crosstalk~\cite{Barends2016,Endo2021}. Understanding and reducing this \emph{control overhead} has therefore become a central problem in NISQ algorithm design and compilation.\\

	Among the most costly circuit primitives are multi-qubit controlled operations, particularly the controlled-$U$ gate. Controlled unitaries are central to many protocols, including phase estimation~\cite{NielsenChuang2010} and amplitude estimation~\cite{Brassard2002}, yet implementing a controlled version of an arbitrary unitary typically requires a full decomposition of $U$ with an additional control line, increasing both depth and hardware complexity, creating noise problems which must be dealt with~\cite{Mitarai2018,Endo2021,Temme2017,Fellous2025}. 
	
	Several strategies have been proposed to mitigate this cost, including circuit reformulations that replace indirect Hadamard-test measurements by direct expectation-value estimators~\cite{mitarai2019,polla2023} and phase-estimation variants avoiding explicit controlled unitaries~\cite{clinton2024}. However, these approaches either rely on structural assumptions about the target unitary or modify the underlying algorithmic framework, and thus do not provide a general deterministic construction of the logical action of $C(U)$ for arbitrary $U$. As an alternative, there are optimization methods which produce shallow controlled-$U$ circuits by adding ancillary qubits and a sequence of controlled gates which varies for each $U$ \cite{Wu2021}.\\
	
	Here we introduce a deterministic construction that reproduces the action of a controlled unitary using only uncontrolled applications of $U$ and a fixed circuit around the unitary. The method employs an ancilla-assisted double-SWAP network: the control qubit acts on two register SWAPs surrounding a single instance of $U$. When a known eigenstate $\ket{e}$ of $U$ with eigenvalue $\lambda=e^{i\phi}$ is supplied, we show analytically that the resulting operation on the ancilla and system registers is equivalent to a controlled-$U$ up to a phase correction $e^{-\frac{i\phi}{2}}R_Z(\phi)$ (a diagonal matrix introducing a relative $-\phi$ phase shift in the control qubit, with the phase appearing in the $\ket{0}$ term). This replaces the costly synthesis of a controlled-$U$ by two controlled-SWAPs and one bare $U$, yielding a constant-depth overhead and requiring a total of $2n+1$ qubits including the control and eigenstate registers. Notably, multi-qubit controlled gates such as the Toffoli (and CCZ) have been demonstrated natively or in single-step implementations in trapped-ion and superconducting platforms~\cite{goel2021,rasmussen2020,liu2025}, further reducing the practical overhead of the controlled-SWAP components required by our scheme. If $\ket{e}$ is an approximate eigenstate, the deviation with respect to the exact evolution is bounded, ensuring robustness to variationally prepared states. The construction thus establishes an exact, deterministic, and experimentally accessible embedding of $C(U)$, enabling controlled-like functionality in shallow-depth form. It opens a route toward resource-efficient implementations of interferometric primitives such as the Hadamard test or variational phase estimation on near-term hardware, while revealing a conceptual link between eigenstate knowledge and control complexity in quantum computation.\\
	
\begin{figure}[t]
		\centering
		\includegraphics[width=0.97\columnwidth]{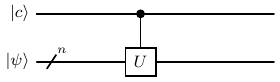}
		(a)\\
		\hfill
		\includegraphics[width=0.97\columnwidth]{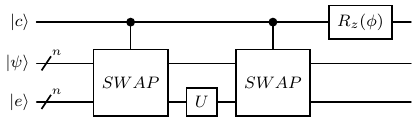}
		(b)
		\caption{(a) Standard controlled-$U$ circuit, where the control qubit directly triggers the unitary $U$.
			(b) Eigenstate-assisted implementation: the ancilla controls two register-SWAP operations surrounding a single, uncontrolled application of $U$ on the auxiliary register. 
			When the auxiliary register is initialized in an eigenstate $|e\rangle$ of $U$, both circuits produce identical transformations on the ancilla--system subspace, up to a local single-qubit phase correction on the ancilla, with an additional overall global phase $e^{i\phi/2}$. 
			Each controlled-SWAP (Fredkin) gate can be further decomposed into a \textsc{CNOT--TOFFOLI--CNOT} sequence, providing a constant-depth, hardware-compatible realization of the scheme.}
		\label{fig1}
	\end{figure}

	\textit{Control-free construction.—}
	We consider a three-register system composed of a control qubit $c$, a target register $S$ holding the input state $\ket{\psi}$, and an auxiliary register $E$ initialized in an eigenstate $\ket{e}$ of $U$ with eigenvalue $\lambda$ ($U\ket{e}=\lambda\ket{e}$, $|\lambda|=1$). 
	The ancilla controls two SWAP operations between $S$ and $E$, with an uncontrolled application of $U$ on the \emph{auxiliary} register in between, as depicted in Fig.~\ref{fig1}(b). Fig.~\ref{fig1}(a) shows the standard controlled-$U$ circuit for comparison.

	Formally, the overall unitary implemented by the circuit in Fig.~\ref{fig1}(b) before phase correction is
	\begin{equation}
		W = \mathrm{CSWAP}_{a,SE}\,(I_S \otimes U_E)\,\mathrm{CSWAP}_{a,SE}.
	\end{equation}
	where $\mathrm{CSWAP}_{a,SE} = P_0 \otimes I_{SE} + P_1 \otimes S_{SE}$ with $P_i = \ket{i}\!\bra{i}$ the projectors on the control and $S_{SE}$ the SWAP between registers $S$ and $E$. 
	Using $S_{SE}(I_S \otimes U_E)S_{SE} = U_S \otimes I_E$, a straightforward calculation yields
	\begin{equation}
		W = P_0 \otimes (I_S \otimes U_E) + P_1 \otimes (U_S \otimes I_E).
		\label{eq1}
	\end{equation}
	
	Acting on an input $\ket{c}_a\ket{\psi}_S\ket{e}_E$ with $\ket{c} = \alpha\ket{0} + \beta\ket{1}$, Eq.~(\ref{eq1}) gives
	\begin{equation}
		\begin{aligned}
			W(\ket{c}\ket{\psi}\ket{e}) 
			&= \alpha\ket{0}\ket{\psi}U\ket{e} + \beta\ket{1}U\ket{\psi}\ket{e} \\
			&= \alpha\lambda\ket{0}\ket{\psi}\ket{e} + \beta\ket{1}U\ket{\psi}\ket{e} \\
			&= \left(\bigl(P_0 \otimes \lambda I + P_1 \otimes U \bigr)\ket{c}\ket{\psi}\right)\ket{e} \\
			&= \bigl(e^{\frac{i\phi}{2}}R_Z(-\phi)\, C(U)\,\ket{c}\ket{\psi}\bigr)\ket{e}.
		\end{aligned}
		\label{eq2}
	\end{equation}
	where $C(U)=P_0\otimes I + P_1\otimes U$ is the conventional controlled-$U$ (with $U$ acting on the target in register $S$, as in Fig.~\ref{fig1}(a)). 
	
	Equation~(\ref{eq2}) shows that, up to a known single-qubit phase correction on the ancilla, the transformation on the ancilla–system subspace is identical to the action of a controlled-$U$. Hence, the controlled operation is recovered \emph{without ever applying a control on $U$ itself.}\\
	
	This establishes a \textit{deterministic eigenstate-assisted embedding} of arbitrary unitaries, requiring only controlled two-register SWAP operations and one bare application of $U$, independent of the internal decomposition of $U$. 
	
	\textit{Implementation and error robustness.—}
	In terms of resource cost, a direct synthesis of the controlled version of an arbitrary $U$ requires decomposing $U$ into one- and two-qubit gates and promoting each to its controlled counterpart, resulting in $\mathcal{O}(n d_U)$ two-qubit operations for an $n$-qubit unitary of depth $d_U$~\cite{Barenco1995,NielsenChuang2010}. For the totally connected qubits of the usual circuit model, there are schemes that use $s\leq n$ ancillary qubits to give an $\mathcal{O}(\log s + d_U\log \frac{n}{s} + d_U)$ depth, with a constant factor of $9$ in the $d_U$ term (see Lemma 6 in \cite{Wu2021}), which we consider the dominant part in the depth. In contrast, the present construction replaces this entire control overhead by two controlled-SWAP (Fredkin) gates and one bare application of $U$, producing a constant-depth additive term independent of the structure of $U$ with $n$ ancillas and the knowledge of an eigenstate of $U$.
	
	Each controlled-SWAP can be decomposed as \textsc{CNOT--Toffoli--CNOT} for each qubit, corresponding to 8 effective CNOTs per qubit in the standard circuit model~\cite{Shende2009}. Thus, the full circuit, comprising two CSWAPs and one instance of $U$, requires at most about $16n$ two-qubit operations, independent of the depth $d_U$ and the internal structure of $U$.
	
	The only additional resource is the auxiliary register holding the eigenstate $\ket{e}$, which can be prepared once and reused across evaluations. For an $n$-qubit unitary, this corresponds to a total of $2n+1$ qubits, including the control, representing a fixed increase in register width that accompanies the reduction in circuit depth. 
	
	The eigenstate register effectively acts as a small quantum memory. Apart from the single bare application of $U$, its only involvement in the circuit is through the two CSWAP operations, which merely permute its position without modifying its internal amplitudes. Consequently, its coherence requirements reduce to maintaining the eigenstate structure of $\lvert e\rangle$. Any residual dynamical phase accumulated between runs contributes only a global phase and leaves the logical action of the scheme unchanged. As a result, control complexity is shifted from gate synthesis to state preparation, with the latter amortized over repeated uses.
	
	The construction relies on the availability of an eigenstate $\ket{e}$ of the unitary $U$, which may at first appear to limit its applicability. However, a wide class of practical unitaries, especially those arising from Hamiltonian simulation or variational models, possess eigenstates that can be efficiently prepared on NISQ devices. Variational quantum eigensolvers (VQE)~\cite{Peruzzo2014,McClean2016} and their extensions~\cite{Larose2019,Kardashin2020,Cerezo2021,Tilly2022,Cerezo2022} can approximate eigenstates of a Hermitian generator $H$ such that $U = e^{-iH}$ using circuits of moderate depth and no controlled operations. Hardware-efficient Ans\"atze~\cite{Kandala2017} and adaptive variational strategies~\cite{Grimsley2019} can significantly reduce circuit depth while achieving high overlaps with target eigenstates in current platforms.

For general unitaries $U$, variational methods based on the SWAP test can approximate eigenstates using shallow circuits~\cite{Garcia2024}. The corresponding eigenvalue can then be extracted using phase-estimation--based algorithms~\cite{Moore2021} or through optimized Hadamard-test schemes~\cite{Wu2021}. Although these procedures still involve controlled unitaries, their implementation cost can remain within the capabilities of current NISQ devices. After this initial eigenvalue-learning stage, a fixed-cost circuit of reduced depth can be employed in larger circuits, effectively shifting the control overhead to a one-time quantum compilation step.

In all the cases, the training stage to find the circuit generating the eigenstate only happens once for each type of controlled unitary. 
	
To quantify robustness against imperfect eigenstate preparation, we can study the fidelity of the generated state under small perturbations.

Let $\ket{c}=\alpha\ket{0}+\beta\ket{1}$ be the control qubit and $\ket{\psi}$ a general $n$-qubit state. We want to reproduce the action of a controlled unitary $$C(U)\ket{c}\ket{\psi}=\alpha\ket{0}\ket{\psi}+\beta\ket{1} U\ket{\psi}$$ with our circuit follower by a phase correction.

We need to compare the ideal output state $$\ket{\phi_i}=\alpha\ket{0}\ket{\psi}\ket{e}+\beta\ket{1}U\ket{\psi}\ket{e}$$ to the output of our circuit when the auxiliary input has an approximation to the actual eigenstate of $U$ such that $$\ket{\tilde{e}}=\sqrt{1-\epsilon}\ket{e}+e^{i\varphi}\sqrt{\epsilon}\ket{\perp}$$ for a real $|\epsilon|\ll 1$,  where $\braket{e|\perp}=0$ and $\bra{e}U\ket{\perp}=0$. The approximated output, before phase correction, is the state
\begin{align}
\ket{\phi_a}&=W\ket{c}\ket{\psi}\ket{\tilde{e}}&\\
&=\sqrt{1-\epsilon}\lambda\alpha\ket{0}\ket{\psi}\ket{e}+e^{i\varphi}\sqrt{\epsilon}\alpha\ket{0}\ket{\psi}U\ket{\perp}&\\
&+\sqrt{1-\epsilon}\beta\ket{1}U\ket{\psi}\ket{e}+e^{i\varphi}\sqrt{\epsilon}\beta\ket{1}U\ket{\psi}\ket{\perp}.&
\end{align}
The fidelity between the exact and the approximated output states, after phase correction, is
$$|\bra{\phi_i}R_Z(\phi)\ket{\phi_a}|^2=|\sqrt{1-\epsilon}(|\alpha|^2+|\beta|^2|)|^2=1-\epsilon.$$

Thus, the deviation from the ideal controlled-$U$ action is comparable to the eigenstate inaccuracy, and the equivalence remains exact in the limit $\epsilon \to 0$.

	This behavior can help in the synthesis of controlled-$U$ from elementary gates, where imperfections in each controlled subgate accumulate with circuit depth, leading to a depth-dependent increase of the overall error~\cite{Preskill2018,Buruaga2025}.
	Our approach therefore converts \emph{coherent control complexity} into a \emph{state-preparation accuracy} requirement, a trade-off naturally suited for variational or hybrid quantum--classical frameworks. 
	If a variational Ansatz with $L$ layers is used to prepare $\ket{e}$, we have at most a depth $L$ and at most $\mathcal{O}(nL)$ two-qubit gates for an $n$-qubit system. Many hardware-efficient Ans\"atze are widely used in VQE implementations and exhibit linear scaling in both depth and parameter count \cite{Kandala2017,Sim2019,Cerezo2021,Tilly2022}. Training happens only once and there are many known efficient algorithms for eigenstate estimation in variational settings.\\

	\textit{Applications and outlook.—}
	A direct application of the control-free embedding is the Hadamard test, which estimates the complex expectation value $\braket{\psi|U|\psi}$. In the standard circuit, the ancilla is initialized in $\ket{+} = (\ket{0} + \ket{1})/\sqrt{2}$, a C(U) gate controlled by the ancilla is applied to the target state and then the ancilla is measured in the $\{\ket{+},\ket{-}\}$ basis (a Hadamard gate followed by measurement in the computational basis). The measured expectation values of the $X$ and $Y$ observables on the ancilla after the control give the real and imaginary parts of $\braket{\psi|U|\psi}$, respectively. They correspond to a Hadamard gate or an $S^\dag$ and a Hadamard gate before a computational basis measurement.
	
	In our control-free version, the controlled-$U$ is replaced by the double-SWAP construction acting on registers $S$ and $E$, with the auxiliary register initialized to the eigenstate $\ket{e}$ satisfying $U\ket{e} = \lambda\ket{e}$. Acting on the total state
		\begin{equation}
			\ket{+}_a\ket{\psi}_S\ket{e}_E = \frac{1}{\sqrt{2}}(\ket{0}\ket{\psi}\ket{e} + \ket{1}\ket{\psi}\ket{e}),
		\end{equation}
		and using Eq.~(2) from above, we obtain
		\begin{align}
			W\ket{+}\ket{\psi}\ket{e}
			&= \frac{1}{\sqrt{2}}
			\Big(\ket{0}\ket{\psi}U\ket{e} + \ket{1}U\ket{\psi}\ket{e}\Big) \notag\\
			&= \frac{1}{\sqrt{2}}
			\Big(\lambda\,\ket{0}\ket{\psi}\ket{e} + \ket{1}U\ket{\psi}\ket{e}\Big).
		\end{align}

	Expanding the full pure-state density operator gives
	\begin{equation}
		\begin{aligned}
			\varrho_{aSE}=|\Psi\rangle\!\langle\Psi|
			&=\tfrac12\Big(
			|0\rangle\!\langle0|\!\otimes\!|\psi\rangle\!\langle\psi|\!\otimes\!|e\rangle\!\langle e|\\
			&+|1\rangle\!\langle1|\!\otimes\!U|\psi\rangle\!\langle\psi|U^\dagger\!\otimes\!|e\rangle\!\langle e|\\
			&+\lambda\,|0\rangle\!\langle1|\!\otimes\!|\psi\rangle\!\langle\psi|U^\dagger\!\otimes\!|e\rangle\!\langle e|\\
			&+\lambda^{*}\,|1\rangle\!\langle0|\!\otimes\!U|\psi\rangle\!\langle\psi|\!\otimes\!|e\rangle\!\langle e|
			\Big).
		\end{aligned}
		\label{eq:rho_full}
	\end{equation}
	After tracing out the system and eigenstate registers $S$ and $E$, the reduced ancilla density matrix is
	\begin{equation}
		\rho_a = \frac{1}{2}
		\begin{pmatrix}
			1 & \lambda\braket{\psi|U^\dagger|\psi} \\
			\lambda^*\braket{\psi|U|\psi} & 1
		\end{pmatrix}.
	\end{equation}
	Therefore, measuring $X$ and $Y$ observables for the ancilla yields
	\begin{equation}
     \label{expected}
		\langle X\rangle = \mathrm{Re}\!\left[\lambda^*\braket{\psi|U|\psi}\right],
		\qquad
		\langle Y\rangle = \mathrm{Im}\!\left[\lambda^*\braket{\psi|U|\psi}\right].
	\end{equation}
	
	\begin{figure}[t]
		\centering
		\includegraphics[width=0.97\columnwidth]{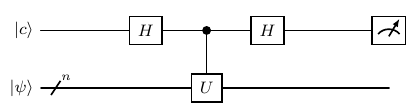}
		(a)\\
		\hfill
		\includegraphics[width=0.97\columnwidth]{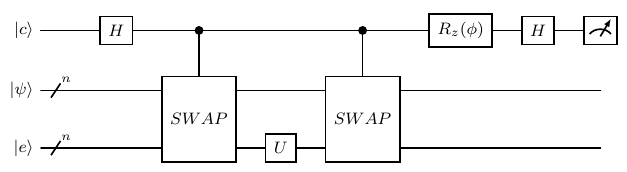}
		(b)
		\caption{(a) Hadamard test with controlled-$U$: ancilla $H$, $C(U)$, ancilla $H$, and a $Z$ measurement (for estimation of the real part).
			(b) Control-free scheme: ancilla $H$, CSWAP--$U$--CSWAP, optional ancilla $R_Z(\phi)$, ancilla $H$, and measurement.
			Both schemes yield identical estimators of $\langle\psi|U|\psi\rangle$, with the control-free scheme producing a known multiplicative phase factor $\lambda$ that can be removed either by the optional $R_Z(\phi)$ gate or by classical post-processing, and an additional overall global phase $e^{i\phi/2}$.
		}
		\label{fig2}
	\end{figure}
	
    A known single-qubit phase correction on the ancilla restores the exact statistics of the conventional Hadamard test. Knowing $\lambda$, we can skip the phase correction and recover the expected values from Eqs. (\ref{expected}) and a quick computation. Hence, \emph{the same expectation values can be extracted without any controlled application of $U$}, at identical sampling cost and with a deterministic outcome. The only additional resource is the auxiliary register $\ket{e}$, which can be reused across experiments as long as its coherence is preserved.

	In quantum machine learning, many kernel and general inner-product estimators rely on variants of the Hadamard test to evaluate overlaps between data-encoding states~\cite{Liu2018,Schuld2020,Xu2022}. These protocols often require controlled implementations of the feature-map unitary $U(\theta)$, substantially increasing circuit depth on NISQ hardware. Control-free alternatives have been explored in specific overlap-estimation settings by reformulating indirect measurements as direct expectation-value estimators~\cite{mitarai2019,polla2023}. Our eigenstate-assisted construction provides a deterministic and constant-depth generalization applicable to arbitrary controlled unitaries.
	
	The broader implication is that \emph{control complexity can be traded for spectral information}: knowledge of an eigenstate of $U$ suffices to simulate control over $U$ with a constant overhead (two register-SWAPs plus one bare $U$) and errors tied solely to the eigenstate fidelity. This shift from gate-synthesis burden to state-preparation accuracy aligns with NISQ capabilities and suggests refined resource accounting where eigenstate knowledge becomes a first-class primitive.  Given that two-qubit entangling gates and single-qubit phase rotations are native operations in leading trapped-ion and superconducting platforms, the SWAP-based structure of our construction can be readily implemented on existing hardware~\cite{goel2021,rasmussen2020}, placing near-term demonstrations—such as Hadamard tests for molecular unitaries or learning kernels—within immediate experimental reach.\\
	
	\textit{Conclusion.—}
	We have introduced a control-free framework for realizing the logical action of a controlled unitary using only uncontrolled applications of $U$ and ancilla-assisted SWAP interactions. By exploiting the existence of an eigenstate of $U$, the circuit effectively transfers control from the level of gate synthesis to the level of state preparation, providing an exact and deterministic alternative to conventional controlled-$U$ implementations. This construction reduces both circuit depth and control overhead while preserving full unitarity, and it achieves a constant cost in terms of two controlled-SWAPs and one bare $U$, independent of the internal decomposition of $U$.
	
	Beyond its immediate practical impact on near-term quantum devices, this result exposes a deeper structural link between \emph{control complexity} and \emph{spectral knowledge}. In essence, having access to an eigenstate of $U$ is sufficient to simulate control over $U$ itself. This insight suggests that controlled operations, traditionally viewed as separate resources, can instead be encoded through eigenstate correlations, potentially redefining how we quantify quantum circuit resources. The idea parallels concepts in resource theory, where structural information replaces operational overhead \cite{howard2017}, and may inspire new paradigms for state-driven computation.
	
	\textit{Acknowledgments} 
	J.C. Garc\'ia Escart\'in has been funded by Junta de Castilla y Le\'on (Consejer\'ia de Educaci\'on/FEDER) Project VA184P24 and the European Union NextGeneration UE/MICIU/Plan de Recuperaci\'on, Transformaci\'on y Resiliencia/Junta de Castilla y Le\'on. Financial support of the Department of Education, Junta de Castilla y León, and FEDER Funds is gratefully acknowledged (Reference: CLU-2023-1-05).

\end{document}